\documentclass{article}

\usepackage{multirow}
\usepackage{setspace} 
\usepackage{authblk}
\usepackage{physics}
\usepackage{graphicx}
\usepackage{dcolumn}
\usepackage{bm}
\usepackage[mathlines]{lineno}
\usepackage{color,soul}

\begin{document}

\title{Methods for removal of unwanted signals from gravity time-series: comparison using linear techniques complemented with analysis of system dynamics}

\author[]{Arthur Valencio\thanks{Corresponding author: a.valencio@abdn.ac.uk}}
\author[]{Celso Grebogi}
\author[]{Murilo S. Baptista}
\affil[]{Institute for Complex Systems and Mathematical Biology, University of Aberdeen, Aberdeen \emph{AB24 3UE}, United Kingdom}

\date{}

\maketitle

\begin{abstract}
The presence of undesirable dominating signals in geophysical experimental data is a challenge in many subfields. One remarkable example is surface gravimetry, where frequencies from Earth tides correspond to time-series fluctuations up to a thousand times larger than the phenomena of major interest, such as hydrological gravity effects or co-seismic gravity changes. This work discusses general methods for removal of unwanted dominating signals by applying them to 8 long-period gravity time-series of the International Geodynamics and Earth Tides Service, equivalent to the acquisition from 8 instruments in 5 locations representative of the network. We compare three different conceptual approaches for tide removal: frequency filtering, physical modelling and data-based modelling. Each approach reveals a different limitation to be considered depending on the intended application. Vestiges of tides remain in the residues for the modelling procedures, whereas the signal was distorted in different ways by the filtering and data-based procedures. The linear techniques employed were power spectral density, spectrogram, cross-correlation and classical harmonics decomposition, while the system dynamics was analysed by state-space reconstruction and estimation of the largest Lyapunov exponent. Although the tides could not be completely eliminated, they were sufficiently reduced to allow observation of geophysical events of interest above the $10 \text{ nm s}^{-2}$ level, exemplified by a hydrology-related event of $60 \text{ nm s}^{-2}$. The implementations adopted for each conceptual approach are general, so that their principles could be applied to other kinds of data affected by undesired signals composed mainly by periodic or quasi-periodic components.
\end{abstract}


\begin{doublespacing}
\section{Introduction}

Many geophysical data present strong periodic or quasi-periodic signals masking the observations of the phenomena of interest. For the case of precise surface gravimetry, phenomena of interest are polar ice cap variations and melting \cite{Sato2006}, hydrological effects including remote assessment of underground water reservoirs \cite{Imanishi2006} forest evapotranspiration rates \cite{VanCamp2016}, co-seismic and post-seismic deformations \cite{Imanishi2004,Soldati1998}, and proposals of gravity-field perturbations before arrival of compressional seismic waves \cite{Montagner2016}. However these effects, typically on the range of $0.1-100 \text{ nm s}^{-2}$, are hindered by gravity tides, with amplitudes of $2000\text{ nm s}^{-2}$.

The device adopted for such gravity applications is the superconducting gravimeter, with precision level smaller than $1\text{nm s}^{-2}$. Its operation is a mathematical equivalent of an ideal spring, built by the employment of a superconducting sphere suspended on a magnetic field produced by persistent currents in a coil. In such device, changes of local gravity induces changes to the equilibrium of the system, which is translated to a relative gravity measurement \cite{Goodkind1999}. A global network of these instruments, the Global Geodynamics Project (GGP), was implemented in the 1990's to address common problems to the gravimetry community \cite{Crossley2009}, and the project is now taken over by the International Geodynamics and Earth Tide Service (IGETS), enabling the access to data from 33 stations (http://isdc.gfz-potsdam.de/igets-data-base). However, the issue of tidal filtering remains a challenge with many possible solution approaches and no full consensus.

The tides observed represent several single frequencies mostly around the diurnal, semi-diurnal and terdiurnal values. Despite progress in the development of more complete tidal tables, the Darwin nomenclature \cite{Darwin1907} for the main tidal modes remains in use for easy identification (\emph{e.g.} diurnal: K1, O1; semidiurnal: M2, S2; terdiurnal: M3, MK3). The highest amplitude tidal signals are able to influence or mask geophysical observations conducted in frequencies below 5 cycles per day ($5.79\cdot10^{-5}\text{ Hz}$), including ground displacement GNSS measurements, ground strain and monitoring of water-body levels. Other systems may have unwanted frequencies different from tides, due to contamination from other sources. Hence general methods are necessary as the first procedure of analysis.

In this paper, we consider three conceptually distinct general methods for removing undesired periodic signals: (i) frequency filtering of the components (known from spectroscopy or model); (ii) physical modelling of the contributing sources to the signal and subtraction from observation; (iii) and data-based modelling. The latter infers from the data the parameters that better describe the unwanted signal and extract the residual. The description of how each method is implemented and applied to gravity tides is described in Sec. \ref{sec:filtermet}. Although special focus is given to the gravity time-series, the principles are general and comprehensive to the preliminary signal analysis of many applications, and the necessary adjustments should be kept to a minimum. In the results (Sec. \ref{sec:results}) it is shown that a FFT-based frequency filtering may appear effective but the artificial removal of information near tidal frequencies and the addition of the Gibbs ringing phenomenon generates undesirable consequences for the observation of geophysical events of interest or other physical applications. A frequency filtering based on a multiband filter also distorts the frequency spectrum, and is shown to be unable to completely remove all tidal components. The physical modelling reduces the tidal oscillations, but tidal peaks remain present in the frequency domain. Time-series analysis based on nonlinear approaches for the state-space reconstruction and the estimation of Lyapunov exponents shows that the nonlinear features of the original time-series is preserved in the residual from the physical modelling. However, comparatively this is the method where the original physical nature of the system is most preserved based on the state-space reconstruction and sensitivity to initial conditions. The data-based method has the best performance in reducing the gravity residuals without appearing to distort the frequency spectrum, though the state-space plot features are less preserved. The residuals from this method enable the observation of hydrology-induced gravity changes, exemplified in Sec. \ref{sec:appl}.

\section{Materials and methods}

\subsection{Selected stations}

\begin{figure}
    \centering
    \includegraphics[width=90mm]{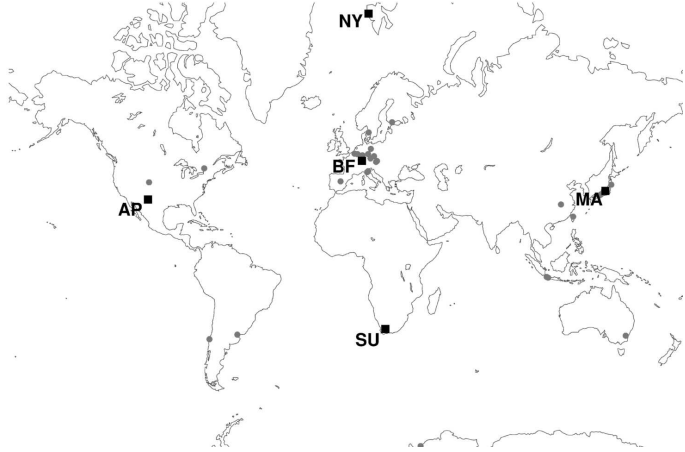}
    \caption{Location of the IGETS superconducting gravimetry stations. The locations of the stations used for this study are presented in black squares. Gray circles represent the other IGETS stations with data currently available to users}
    \label{fig:stationmap}
\end{figure}

\begin{figure}
    \centering
    \includegraphics[width=7.0cm]{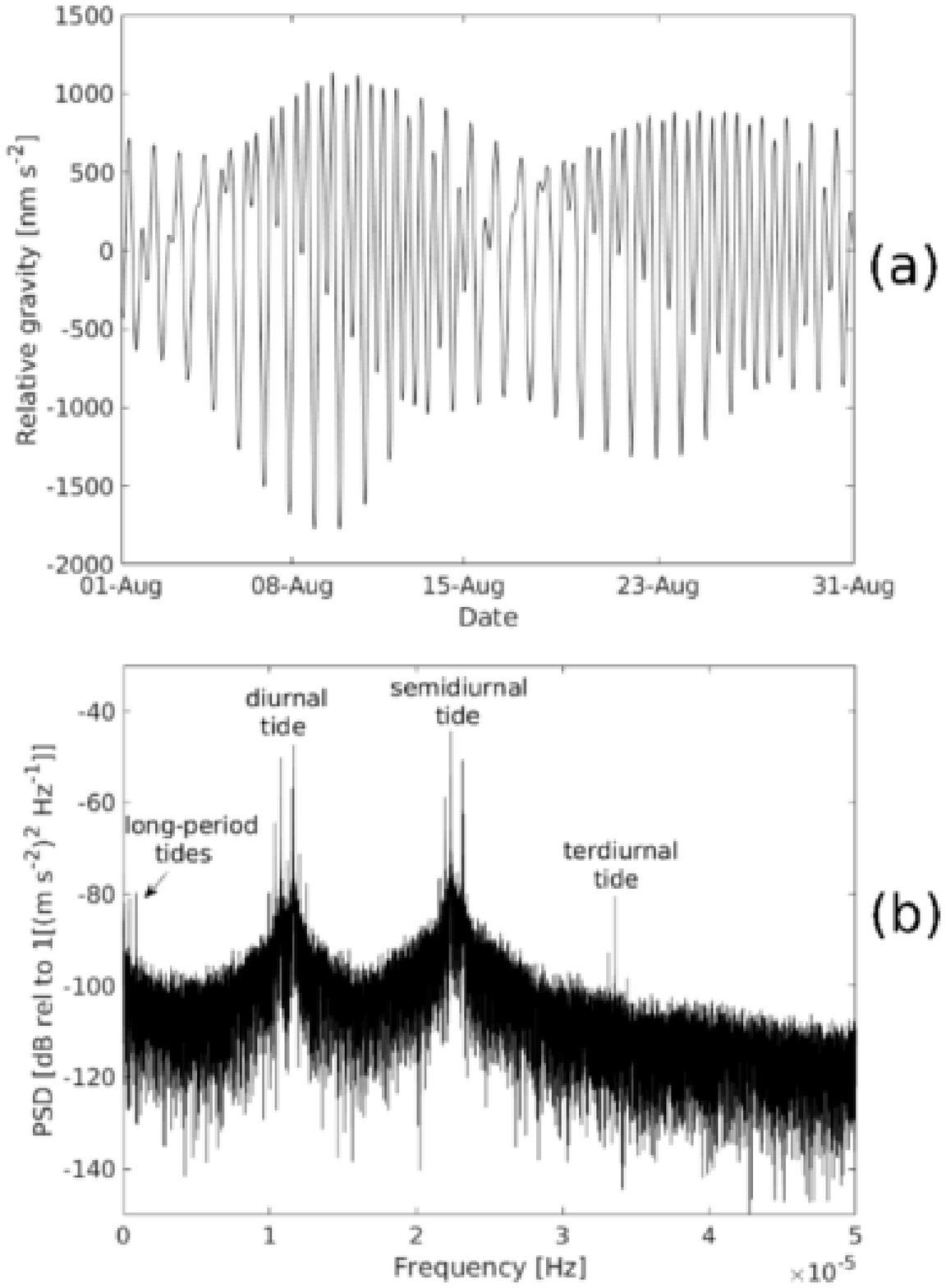}
    \caption{Example of gravity signal from the SU3 instrument, in South Africa. Top panel: 1 month sample (August 2010) of the relative gravity measurements. Bottom panel: Power spectrum density of the entire time-series. Diurnal, semi-diurnal and terdiurnal tides are particularly evident, as indicated. Long-period tides can also be observed, both as a modulation on the time-series and as small peaks in the very low frequencies in the bottom panel.}
    \label{fig:eg-su3}
\end{figure}

For this study, Superconducting Gravimeters were selected located on Sutherland in South Africa (SU, 3 instruments), Schiltach/Black Forest in Germany (BF, 2 instruments), Ny-\AA{}lesund in Svalbard island, Norway (NY, 1 instrument), Matsushiro in Japan (MA, 1 instrument), and Apache Point in New Mexico, USA (AP, 1 instrument), which distributed according to Fig. \ref{fig:stationmap}. The data period of the time-series used and the type of each instrument that generated it are detailed in Table \ref{tb:instruments}. The chosen instruments are a representative sample of the IGETS network, including different latitudes, different site conditions (\emph{e.g.} continentality, vegetation, climate conditions, etc), and different generations of gravimeter instruments. For this analysis it was used 1-min sampling time-series data. Example of time-series and frequency spectrum are shown in Fig. \ref{fig:eg-su3}. 

\begin{table}
\centering
\caption{Instruments and period of the datasets used in this study}
\resizebox{80mm}{!}{
\begin{tabular}{lll}
Instr. & Gravimeter type       & Data period               \\
AP         & Observatory SG (3rd gen)              & 01/Jan/2009 to 31/Dec/2015\\
BF1        & Observatory SG (3rd gen)              & 01/Oct/2009 to 31/Dec/2015\\
BF2        & Observatory SG (3rd gen) 			   & 01/Oct/2009 to 31/Dec/2015\\
MA         & ``Tidal" SG (1st gen)                  & 01/Jul/1997 to 30/Jun/2008\\
NY         & Compact SG (2nd gen)                  & 20/Sep/1999 to 31/Jan/2012\\
SU1        & Dual-sphere SG (2nd gen)     		   & 27/Mar/2000 to 31/Dec/2015\\
SU2        & Dual-sphere SG (2nd gen)              & 30/Sep/2000 to 31/Dec/2015\\
SU3        & Observatory SG (3rd gen)      		   & 01/Sep/2008 to 31/Dec/2015\\
\end{tabular}}
\label{tb:instruments}
\end{table}

\subsection{Tidal removal methods}\label{sec:filtermet}

\subsubsection{Pre-processing}

Although local operators provide the IGETS users with data corrected for spikes and clippings related to the helium refill procedure or to strong motions in the location, the time-series still contain data gaps, offsets and instrument linear trends, which must be considered prior to the application of the tidal removal methods. Also the contribution of the atmospheric mass to the gravity signal must be accounted. The three classes of methods considered for tidal removal, frequency filtering, physical modelling and data-based modelling, have different demands of pre-processing. In the case of frequency filtering, data gaps cannot be present, hence missing intervals, which can be as large as many months, must be temporarily replaced with a synthetic gravity signal. Regardless of that, 90\% of the local atmospheric contribution to gravity can be removed by considering a linear proportion between the atmospheric gravity variation ($\delta g_{atm}$) and the air-pressure change ($\delta p$), with the proportionality constant (atmospheric admittance) being $\alpha=\delta g_{atm}/\delta p =-3.56 \text{ [}\text{nm s}^{-2} \text{mbar}^{-1}\text{]}$ \cite{Merrian1992}, which is considered a sufficient atmospheric correction for this method. The physical modelling and data-based modelling procedures do not require filling the gaps in the time-series, and they use a more complete analysis for atmospheric contribution, thus being sufficient to only correct for offsets and instrument linear trends on the pre-processing. This was performed with a semi-automatic implementation of the remove-restore procedure \cite{Hinderer2015} on Matlab.

\subsubsection{Frequency filtering}

This filtering method consists of simply removing the undesirable tidal frequencies. It is mainly adopted as a preliminary analysis of the spectrum or for the investigation of low-frequency seismic modes known to be out-of-resonance with tides. The tidal frequencies were obtained from the Tamura \cite{Tamura1987} tables, containing 1200 constituent waves. These frequencies may be removed from the original gravity signal either in the time-domain or frequency-domain. For the time-domain it was applied a multiband finite-impulse response (FIR) filter with zero-phase distortion, and for the frequency domain it was adopted the classical procedure of analysing the Fourier spectrum, removing the selected frequencies (setting them to zero in the Fourier domain), and reconstructing the time-series by inverse fast Fourier transform (FFT filtering). Original gaps were reintroduced in the residuals, and extra gap margins of 1-week before and after the original gaps were included to remove artificial ringing effects (Gibbs phenomenon). It must be stressed that these procedures for obtaining the gravity residuals are not suitable for all applications. In particular applications involving signals close to resonance with tides or investigation of very broad-band events should not adopt this method. It is also not recommended for studies that require the quantification of informational measures in the time-series, such as studies of causality, since this method artificially removes information from the data. Although Zetler\cite{Zetler1964} arguments that finite-numbered strong periodicities should not be analysed with FFT or removed with frequency filtering procedures, this technique remained common practice to remove tides until much later \cite{Walters1981}, and may still be used in other fields.

\subsubsection{Physical modelling}

This method is based on modelling all known tidal contributing sources (physical events), resulting in a theoretical prediction. Such prediction is then subtracted from the observed signal so to obtain the gravity residuals, which would only contain the events of interest, for example, co-seismic changes. Therefore it is required to properly select the other phenomena that produces gravity changes, and to consider how they are described. For this study it is considered the following effects: solid Earth tides, ocean tidal loading, atmospheric gravity contribution, ocean non-tidal loading, hydrology loading and polar tides.

The solid Earth tides, also referred as body tides, are the direct effects of the gravitational pull of astronomical objects over either a homogeneous or layered Earth model, resulting in ground displacements in the order of tenths of cm and local gravity changes in the order of $\mu\text{m s}^{-2}$. This is calculated through the following tide generating potential \cite{Agnew2007}
\begin{equation}
V_{SE}(\boldsymbol{r};t)=g_{e}\Re\left[\sum_{n=2}^\infty \sum_{m=1}^n c_{nm}^*(t)Y_{lm}(\theta,\phi) \right]\label{eq:tidepot}
\end{equation}
, with ($r$,$\theta$,$\phi$) specifying the location (radial distance, co-latitude, longitude), $g_e$ the mean gravity of the Earth at the equator, $Y$ the spherical harmonics, and $c_{lm}(t)$ the complex coefficients calculated from the attraction of the astronomical bodies, which are typically computed from an harmonic expansion. To such Earth model it is then subsequently added the contribution of the oceans, which changes gravity due to the direct mass movement but also deforms back the ground due to the significant weight of water being periodically redistributed. This effect is the ocean tidal loading, computed from a tidal generating potential given by \cite{Farrell1973}
\begin{equation}
V_{OL}(\boldsymbol{r};t)=\rho \iint_{\mathrm{ocean}}G(\abs{\boldsymbol{r}-\boldsymbol{r'}})H(\boldsymbol{r'})dS\label{eq:tideoce1}
\end{equation}
, with $\rho$ the density of water, $H(\boldsymbol{r}')$ the tide at the ocean in the location $r'$, and $G(\abs{\boldsymbol{r}-\boldsymbol{r}'})$ the Green's function for the distance, which appears as solutions for the elastic and Poisson equations of a layered Earth. The gravity tide is obtained from the generating potentials by taking the derivative in the radial direction. The work of Farrell \cite{Farrell1972} describes how the tide generating potentials can also be applied to measurements of tidal displacements, tilts, or strain, and, in addition, exemplifies how the Green's function used can be obtained from first principles for a layered Earth model. Both the ocean tidal loading and the solid Earth tides were computed for the selected stations using the software ATLANTIDA3.1\_2014 \cite{Spiridonov2015}, with the assumptions of tidal periodicities following the Tamura \cite{Tamura1987} tables, layered Earth model IASP91, and ocean model FES2012.

Mass redistribution in the atmosphere also causes significant gravity variations, up to the order of $100\text{nm s}^{-2}$. Analogously to what occurs with the ocean tidal loading, the atmospheric mass redistribution leads to fluctuations on the surface of ground and oceans, in particular with the oceans responding as an inverted barometer for periods larger than one week. Although the atmospheric contribution is dominated by the local admittance, based on reading the air pressure at the station, a full description involves computing local and non-local air mass displacements. 
There are two services providing numerical results of atmospheric gravity based on finite element models: Atmospheric Loading service, provided by EOST/University of Strasbourg (http://loading.u-strasbourg.fr/sg\_atmos.php), and ATMACS, provided by BKG (http://atmacs.bkg.bund.de). For this study it has been adopted the first, selecting the atmospheric data provided by ECMWF ERA-Interim reanalysis (http://www.ecmwf.int/en/research/climate-reanalysis/era-interim), which is able to cover the entire gravity time-series period for all stations. The data, however, is sampled in 6h, and interpolation is necessary.

The ocean non-tidal loading refers to other changes in the water mass due to circulation of currents and wind forcing. These lead to gravity changes that can be larger than ocean tidal loading. Theoretical predictions are calculated similarly to the atmospheric loading, but using the ocean bottom pressure data from the ECCO2 model (http://ecco2.jpl.nasa.gov/). Similar procedure is adopted for the calculation of hydrology loading contributions, due to soil moisture changes, where again it was selected weather data from ERA-Interim reanalysis. Services providing the numerical results using such models for all stations are available from EOST/University of Strasbourg (http://loading.u-strasbg.fr/sg\_ocean.php, http://loading.u-strasbg.fr/sg\_hydro.php). 

Finally, the polar tides are a result of the Chandler wobble, small variation of Earth’s axis of rotation. Using the Earth Orientation data EOPC04 from the International Earth Rotation Service (ftp://hpiers.obspr.fr/iers/eop/eopc04), the polar tides can be calculated as $\delta g_{polar} =-39 \times 10^6 \sin 2\theta (m_1 \cos \phi + m_2 \sin \phi)$ [nm s$^{-2}$], with ($\theta$, $\phi$) again the co-latitude and longitude of the station, and ($m_1$,$m_2$) the equivalent ($x$,$y$) polar motion amplitudes converted to radians \cite{Wahr1985a}.
 
\begin{figure}
    \centering
    \includegraphics[width=10cm]{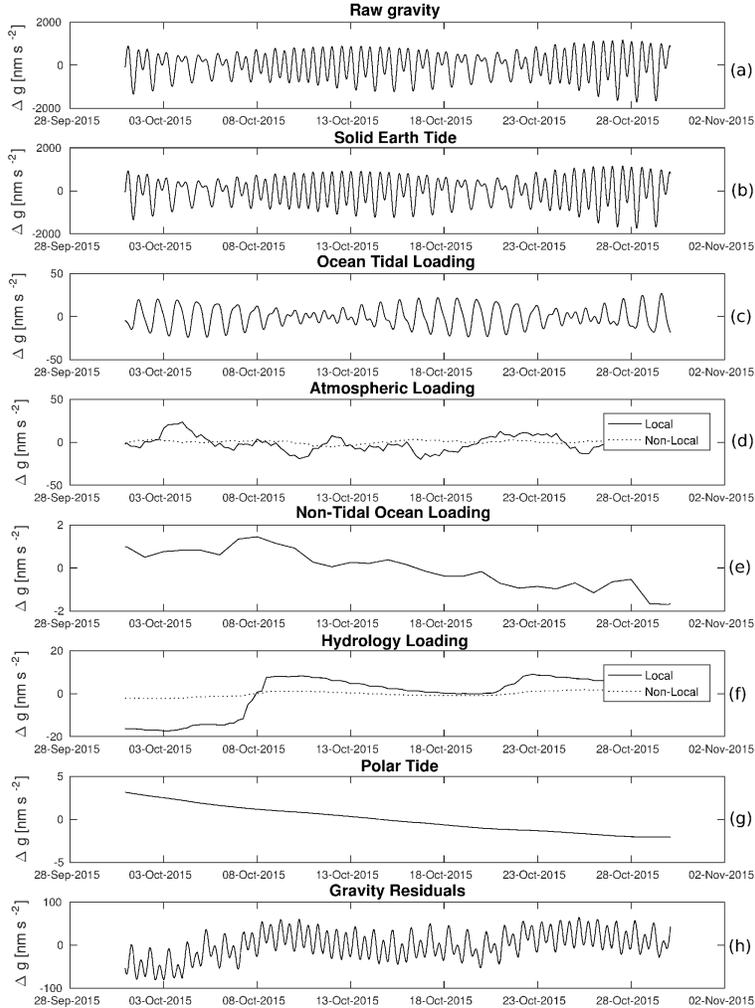}
    \caption{Example of physical modelling applied to 1-month data sample (October 2015) of AP instrument (New Mexico). The gravity residuals (h) are obtained by subtracting the measured relative gravity (a) by all the following simulated contributions: solid Earth tide (b), ocean tidal loading (c), atmospheric loading (d), non-tidal ocean loading (e), hydrology loading (f) and polar tide (g). In this sample, there was a misfitting of the amplitudes of the theoretical semidiurnal tides with the observations, so oscillations are still present in the gravity residuals  (h), albeit with smaller amplitude than in the original time-series (a).}
    \label{fig:eg-physmodel}
\end{figure}

Figure \ref{fig:eg-physmodel} exemplifies the scale from each contribution and procedure for tidal signal removal. A detailed review of these processes can be found in Crossley et al. \cite{Crossley2013}, Hinderer et al.\cite{Hinderer2015} and Boy and Hinderer \cite{Boy2006}. The input parameters for physical modelling refers only to the station location and local/global conditions, and no \emph{a priori} information of the gravity time-series is used. Advantages of this method are the maintenance of information produced on all frequency bands provided by the device, and the possibility of clearly defining the physical origin of any given contribution, including control to maintain aspects of interest according to the application. For example, it is of interest to maintain hydrology loading when the objective of the research is to investigate the gravity response to rainfall, evapotranspiration and aquifer recharge, but still all other contributions should be removed. The limitation in the method is that misfitting in the models introduces undesirable fluctuations to the residuals. Figure \ref{fig:eg-physmodel} (h) makes this evident by the oscillatory pattern with semi-diurnal frequencies. Spectral analysis identifies these oscillatory frequencies in the residuals as corresponding to the tidal modes M2, S2 and K2. Although not adopted in this study, the recently developed software mGlobe\cite{Mikolaj2016} performs similar operations with little more input required from the user than the location of interest. 

\subsubsection{Data-based modelling}

In the previous method the theoretical prediction of tidal mode amplitudes from solid Earth tides and ocean tidal loading was based on a tide generating potential model using only information of location of the instrument and initial time of the dataset as inputs. The tidal modes that remained in the residuals of Fig. \ref{fig:eg-physmodel}(h), however, are strongly associated with these two physical origins, meaning that the theoretical model of these tidal components does not completely fit with the observations at the station. A way of overcoming this is to analyse the regularities in the gravity data itself, extracting from the data the suitable tidal coefficients to the station location. The usual approach is the classical harmonic or least-square fitting, based on defining amplitudes and phases to sinusoids of tabled tidal frequencies, in such a way that the sum of the square of the residual values is minimised. Currently, there is still a lack in the literature about alternative data-based methods, such as the use of Artificial Neural Networks, leading to prediction of gravity tides with the same precision. For compatibility, the Tamura \cite{Tamura1987} tables are again adopted to provide the tidal frequencies for the least-square fitting. Recent tools developed for oceanography, such as UTide \cite{Codiga2011}, have included a series of optimisations, being able to account for data gaps and long time-series without issues, and can be easily adapted to gravity data. The table of data-based tidal constituents is then used to reconstruct only the tidal part of the time-series, and the difference with the observation provides the residue. We have adopted for this analysis the software UTide with the option to use only the Ordinary Least-Squares method. Two issues arise with this method. The first, highlighted by Kantz and Schreiber \cite{Kantz1997}, is that misfitting can occur in the presence of non-white noise, and the assumed background (physical) noise profile of gravity time-series exhibits redness (\emph{i.e.} varies with frequency approximately on $f^{-2}$). The second issue consists in the fact that noise factors or other geophysical events of potential interest might have components near tidal frequencies, and the procedure might mistakenly consider these as part of the contribution of the tidal constituent, hence a type of overfitting. Developments such as the ETERNA \cite{Wenzel1996}, VAV \cite{Venedikov2001}, and BAYTAP \cite{Tamura2008} deal with these aspects in a number of ways, including bandpassing around the tidal frequencies, \emph{i.e.} effectively making an hybrid of data-based and frequency filtering approaches, and additional use of ARMA models to separate the noise and the periodic parts. However these implementations are specifically designed to Earth tides and can require substantial modifications if applied to removal of dominating signal of other geophysical nature, contrary to the more general methods we discuss here. A comparison of performance between these three recent developments was done by Dierks and Neumeyer \cite{Dierks2002}.

\section{Results and discussions}\label{sec:results}

\begin{figure}
    \centering
    \includegraphics[height=18cm]{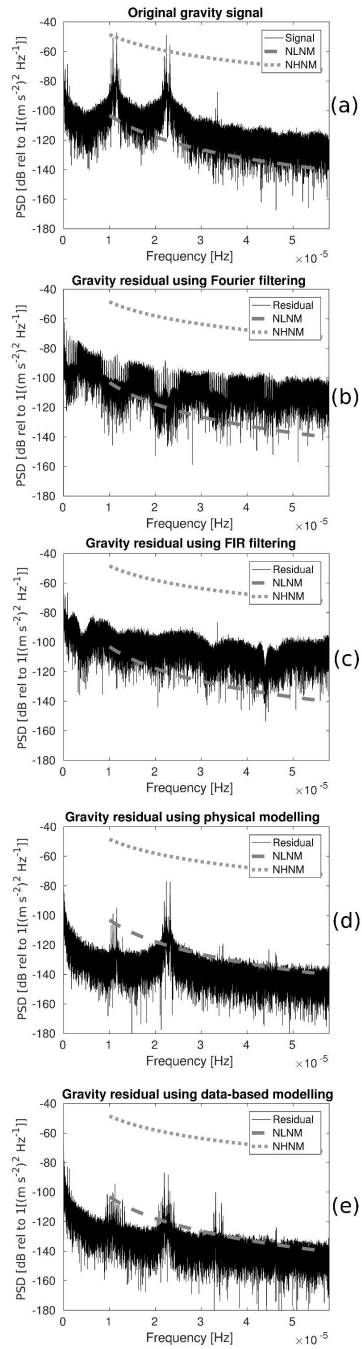}
    \caption{Lomb-Scragle power spectral density of the original gravity signal (a) and gravity residuals (b-e) from BF1 instrument, Schiltach/Black Forest, Germany, as obtained from the different methods of tide removal: FFT (b), FIR filtering (c), physical modelling (d), and data-based modelling using least squares (e). For reference, the New Earth Low Noise Model (NLNM) and the New Earth High Noise Model (NHNM) are indicated. Results for other stations are provided in the Supplementary Material.}
    \label{fig:psd}
\end{figure}

Once the residuals from each method have been obtained, it is of interest to observe if the tides were eliminated by analysing the frequency spectra. Fig. \ref{fig:psd} shows the Lomb-Scragle power spectral density of the residuals for the BF1 instrument, in Schiltach/Black Forest, Germany. This type of power spectrum calculation, is preferred over other types of periodogram for its direct applicability to data with large gaps \cite{Scragle1982}, such as the gravity time-series, and is calculated by
\begin{multline}
P(f)=\frac{1}{2\sigma^2}\Bigg\{\frac{\left[\sum_{i=1}^{N}x_i -\bar{x}\cos{(2\pi f(t_i-\tau))}\right]^2}{\sum_{i=1}^{N} \cos ^2 (2\pi f(t_i-\tau))}+ \frac{\left[\sum_{i=1}^{N}x_i -\bar{x}\sin{(2\pi f(t_i-\tau))}\right]^2}{\sum_{i=1}^{N} \sin ^2 (2\pi f(t_i-\tau))} \Bigg\}
\label{eq:lomb}
\end{multline}.

In the expression, $x_i$ the data points at times $t_i$, and the average and variance given by $\bar{x}$ and $\sigma^2$; the constant $\tau$ is only a time offset that ensures time-invariance during computation. Figures \ref{fig:psd} (d) and (e) show that the diurnal, semidiurnal and terdiurnal tides are still present after filtering with the physical modelling and data-based modelling methods. The amplitude of the semidiurnal tides are slightly larger in the physical modelling case, whereas the data-based modelling reveals larger terdiurnal constituents. However in both cases the highest tidal peaks were reduced below the New Earth High Noise Model (NHMN) reference line, and some tidal constituents, especially the diurnal (around $1.16\cdot10^{-5}\text{ Hz}$), went also below the New Earth Low Noise Model (NHLN) \cite{Peterson1993}. These models provide a limit reference of the spectra of a non-seismic background noise, estimated empirically from IRIS/IDA network of broadband seismometers (but a very seismically quiet site may, in special circumstances, have background noise below NHLN). Frequency filtering (Figs. \ref{fig:psd} (b) and (c)) were able to eliminate the diurnal and semidiurnal tides, as well as the FIR filtering. The effect of FFT filtering in deleting the tidal frequencies is evident in Fig. \ref{fig:psd} (b), with considerable frequency gaps where information is lost. The FIR filtering proved more adequate, once it was implemented to strongly damp the tidal frequencies instead of deleting them. However, due to limitations of design (constrained by the highest order possible to obtain and apply to data), it produced artificial distortions in regions around $0.5\cdot10^{-5}\text{ Hz}$, $3.2\cdot10^{-5}\text{ Hz}$ and $4.4\cdot10^{-5}\text{ Hz}$, while a terdiurnal tide (M3) remained present $3.4\cdot10^{-5}\text{ Hz}$. The more precise modelling of the atmospheric contribution adopted in the physical and data-based modelling have significantly reduced the power of the residuals over the whole range of frequencies plotted. That is revealed by the drop in the base (noise) level in Fig. \ref{fig:psd} (d) and (e) compared to (b) and (c). Due to this, the quaterdiurnal tides, which typically are not observable for their small amplitude compared to background noise, expressed visible peaks at $4.6\cdot10^{-5}\text{ Hz}$ in Figs. \ref{fig:psd} (d) and (e). The spectral results for other stations are similar, with few specificities relating to site conditions; the plots are available in the Supplementary Material.

As the power spectrum indicates that tidal components remain in the residuals, the amplitude levels of the tidal constituents in the residuals were calculated from classical harmonic analysis, with the main observed modes shown in Table \ref{tb:amplitude}. For the Fourier filtering method (FFT), the results confirm that the main peaks were largely eliminated, with largest modes appearing being consistent with noise. The finite-impulse response (FIR) filtering, though, consistently was not able to filter the terdiurnal tides, with the main mode (M3) appearing to even have a small gain, which reveals a issue with the filter design. The physical modelling (PM) was able to eliminate the terdiurnal components, but it remained with noticeable amplitudes in the important diurnal (K1,O1,S1) and semidiurnal (M2,S2,K2) modes, despite reductions above 90\% in the levels of the greatest peaks. Except from rare occasions, possibly related to instrument site conditions, the least-squares data-based modelling (LS) was able to reduce the main peaks to levels below 2 $\text{nm s}^{-2}$, with the largest peak typically not being the original largest mode but a neighbour.

\begin{table*}
\begin{minipage}{130mm}
\centering
\caption{Amplitudes of the main diurnal, semidiurnal and terdiurnal tidal components of original gravity series (column 3) and residuals [$\text{nm s}^{-2}$] (columns 4-7), for each station (column 1). Confidence levels and tidal group are indicated in parentheses. Numbers in bold show high amplitude tidal peaks remaining in the residuals.}
\resizebox{\textwidth}{!}{
\begin{tabular}{lllllll}
Inst. & Tidal component & Original series & FFT residual & FIR residual & PM residual & LS residual \\ \hline
\multirow{3}{*}{AP}  
& Largest diurnal &457.52(0.06)(K1)&0.8(0.3)(S1)& 0.7(0.1)(Q1)& \textbf{13.81(0.05)(K1)}& 0.53(0.05)(S1)\\
& Largest semidiurnal &616.3(0.2)(M2)&0.3(0.1)(MKS2)& 0.3(0.4)(M2)& \textbf{19.4(0.2)(K2)}& 0.6(0.1)(S2)\\
& Largest terdiurnal &9.42(0.05)(M3)&0.43(0.09)(SK3)&\textbf{9.84(0.05)(M3)}& 0.12(0.01)(MK3)& 0.14(0.05)(SK3)\\\hline
\multirow{3}{*}{BF1} 
& Largest diurnal & 493.37(0.08)(K1)& 1.3(0.4)(BET1)& 0.15(0.06)(S1)& \textbf{2.09(0.03)(K1)}& 0.45(0.08)(S1)\\
& Largest semidiurnal & 394.0(0.1)(M2)& 0.8(0.2)(NU2)& 0.04(0.02)(H2)& \textbf{15.9(0.1)(M2)}& 1.0(0.1)(S2)\\
& Largest terdiurnal & 4.60(0.02)(M3)& 0.3(0.2)(M3)& \textbf{4.88(0.03)(M3)}& 0.05(0.01)(MK3)& 0.16(0.02)(MO3)\\\hline
\multirow{3}{*}{BF2} 
& Largest diurnal & 493.37(0.08)(K1)& 1.0(0.4)(BET1)& 0.275(0.08)(2Q1)& \textbf{2.24(0.02)(K1)}& 0.52(0.08)(S1)\\
& Largest semidiurnal & 394.1(0.1)(M2)& 0.6(0.2)(NU2)& 0.07(0.02)(MU2)&\textbf{15.9(0.1)(M2)}& 1.1(0.1)(S2)\\
& Largest terdiurnal & 4.63(0.03)(M3)& 0.2(0.1)(MO3)& \textbf{4.89(0.03)(M3)}& 0.07(0.01)(MK3)& 0.16(0.03)(SK3)\\\hline
\multirow{3}{*}{MA}
& Largest diurnal & 494.4(0.1)(K1)& 0.04(0.03)(TAU1)& 0.5(0.4)(2Q1)& \textbf{18.24(0.09)(K1)} & \textbf{3.3(0.1)(S1)} \\
& Largest semidiurnal & 577.8(0.3)(M2)& 0.11(0.03)(NU2) & 0.3(0.2)(GAM2) &\textbf{16.2(0.3)(M2)} & 1.0(0.3)(S2)\\
& Largest terdiurnal & 8.46(0.09)(M3)& 0.009(0.007)(SK3)& \textbf{8.69(0.09)(M3)}& 0.26(0.05)(SK3) & 0.31(0.08)(SK3)\\\hline
\multirow{3}{*}{NY}
& Largest diurnal & 189.6(0.01)(K1)&0.14(0.03)(S1) & \textbf{3(1)(S1)}& \textbf{4.22(0.08)(O1)}& 0.22(0.09)(S1)\\
& Largest semidiurnal & 21.59(0.05)(M2)& 0.02(0.01) (MU2)&0.6(0.6)(S2)& \textbf{6.32(0.05)(M2)}& 0.18(0.04)(S2)\\
& Largest terdiurnal & 0.20(0.02)(MO3)& 0.01(0.01)(MK3)& 0.2(0.2)(SO3)& 0.24(0.02)(MO3)& 0.02(0.02)(SK3)\\\hline
\multirow{3}{*}{SU1} 
& Largest diurnal &448.27(0.05)(K1) & 0.04(0.02)(S1) & 0.2(0.3)(S1)&\textbf{3.85(0.03)(S1)} & 1.43(0.05)(S1)\\
& Largest semidiurnal & 619.8(0.2)(M2)& 0.05(0.04)(H1)& 0.1(0.1)(L2)& \textbf{54.7(0.3)(M2)}& \textbf{2.0(0.2)(S2)} \\
& Largest terdiurnal & 9.28(0.05)(M3)& 0.005(0.007)(MK3) & \textbf{9.77(0.06)(M3)} &0.277(0.007)(M3) & 0.29(0.05)(SK3)\\\hline
\multirow{3}{*}{SU2} 
& Largest diurnal & 448.30(0.05)(K1)& 0.03(0.02)(O1)& 0.2(0.2)(S1)& \textbf{3.85(0.02)(S1)} & 1.43(0.05)(S1) \\
& Largest semidiurnal & 619.9(0.2)(M2)& 0.02(0.02)(MSN2)& 0.14(0.09)(M2)& \textbf{54.8(0.02)(M2)}& \textbf{2.0(0.2)(S2)}\\
& Largest terdiurnal & 9.29(0.05)(M3)&0.001(0.004)(OQ2)& \textbf{9.78(0.09)(M3)}& 0.282(0.007)(M3)& 0.29(0.06)(SK3)\\\hline
\multirow{3}{*}{SU3} 
& Largest diurnal & 448.30(0.09)(K1)& 0.5(0.1)(BET1)& 0.4(0.6)(NO1)& \textbf{3.72(0.05)(S1)}&1.50(0.09)(S1) \\
& Largest semidiurnal & 619.9(0.2)(M2)&0.5(0.2)(NU2) & 0.4(0.3)(S2)& \textbf{52.9(0.3)(M2)}& \textbf{2.0(0.2)(S2)}\\
& Largest terdiurnal & 9.25(0.07)(M3)& 0.03(0.02)(SK3)& \textbf{9.8(0.2)(M3})&0.28(0.01)(M3)& 0.29(0.08)(SK3)\\
\end{tabular}}
\label{tb:amplitude}
\end{minipage}
\end{table*}

\begin{figure}
    \centering
    \includegraphics[height=16.5cm]{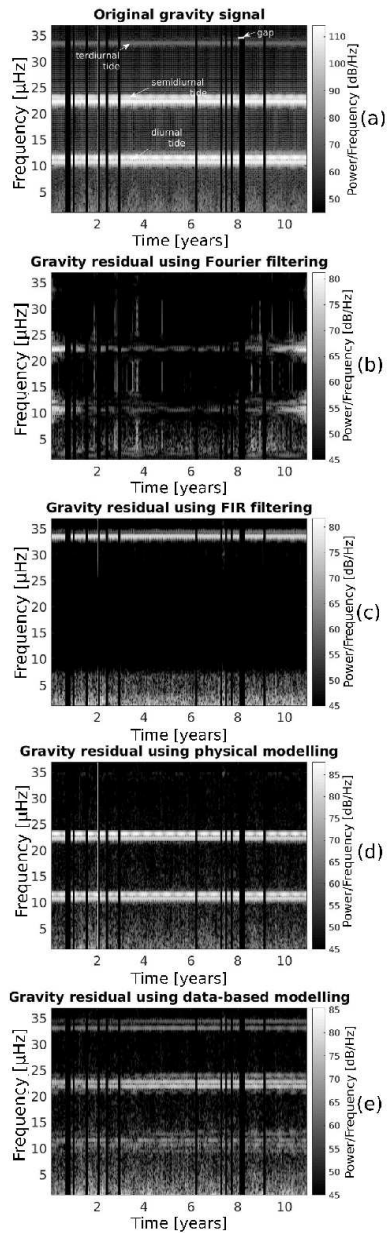}
    \caption{Spectrogram in the range of the main tidal frequencies of the original signal (a) and the residuals (b-e) for the MA instrument, Matsushiro, Japan. The frequency areas of higher energy present in the residuals coincide with the main tidal regions identified in the original gravity signal. Also, these areas of higher energy remain almost constant throughout the whole time-series, as expected from tides. If the cause for the higher energy signals were events localised in time, such as seismic phenomena, the spectrogram would show the energy peaks with well defined start and end times, which is not observed. The pattern is analogous for the other stations. Vertical dark regions are data gaps.}
    \label{fig:spectrogram}
\end{figure}

\begin{figure}
    \centering
    \includegraphics[height=17cm]{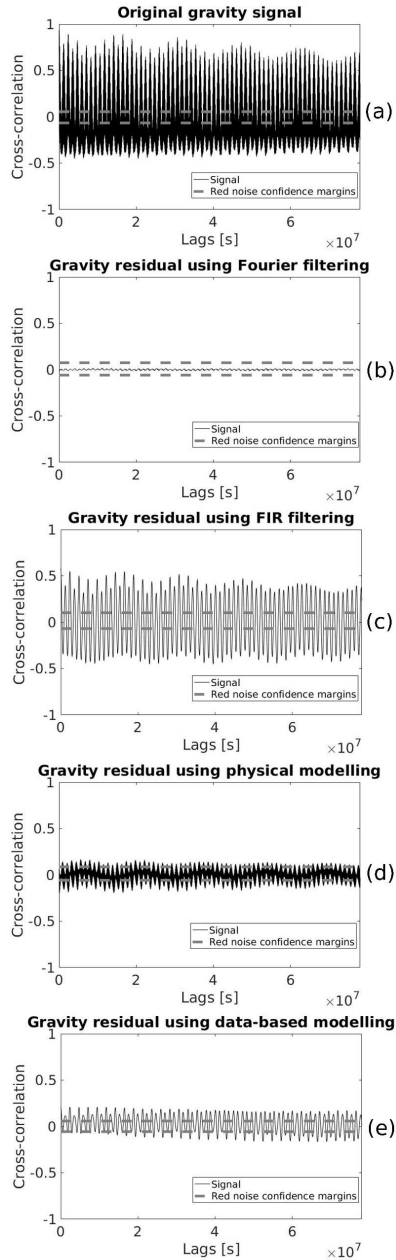}
    \caption{Cross-correlation of original signal (a) and residuals (b-e) with tides on NY instrument, Svalbard island (Norway). Dashed gray lines show confidence margins from correlation between an equal amplitude red noise with the local tides. The maxima of correlation are significantly reduced in comparison with the original signal, particularly with the residuals from FFT filtering (b), physical modelling (d) and data-based modelling (e) at comparable or lower levels than the correlation from a red noise model. FIR filtering was the only case where correlation with tides remained high for this station.}
    \label{fig:tidecorrel}
\end{figure}

A further inquiry is whether the energy in the observed peaks components change in time or remain constant. If it changes, the observed peak in the spectrum might be the result of a temporary effect, of non-tidal origin. In this case, the spectrogram would reveal the energy release to be time-confined. The spectrogram of the residuals (Fig. \ref{fig:spectrogram}) shows, however, uniformity in these frequency contributions along the time, as indicated by the arrows (vertical dark regions are due to time-series gaps). Constant lines at specified frequencies (i.e. horizontal lines crossing the plot) are indications of a particular oscillatory mode being present at all times -- a tide. In a pure gravity residual, these should not appear, instead revealing a more diffuse structure when the time-series is observed as a whole. Reinforcing the argument that tides remain present in the residuals, the cross-correlation between the residuals and the theoretical tides (Fig. \ref{fig:tidecorrel}(d)) is particularly high at all time lags with the physical modelling residuals (d). However, the residuals from other methods had low correlation with theoretical tides [Fig. \ref{fig:tidecorrel} (b), (c) and (e)], below the comparative margins from a red-noise model for gravity residuals. These correlation results reinforce the observation from Table \ref{tb:amplitude} that the physical modelling residuals remain with significant components of the main tides, whereas the residuals from the other methods removes the main tides but maintain the smaller ones. The data-based method produced the best results for tidal filtering without canceling other frequency information, with observed oscillations in the residual time-series typically lower than 100 $\text{nm s}^{-2}$ daily amplitude in the time domain.

Checking for consistency between the residuals from different methods, it is observed that the correlation between the FFT and the FIR residuals is low for all time-series (observed absolute value of Pearson coefficient below 0.03 in all time-series), while it is typically between residuals from physical and data-based modelling (Pearson coefficient above 0.1 for most time-series, reaching up to 0.32 for Apache Point). Exceptions are the time-series from Matsushiro (MA), Japan, and two instruments in Sutherland (SU), South Africa, which also showed low correlations between the residuals from modelling. For MA the local seasonal effects, particularly from atmospheric and ocean events, provides a difficult scenario for the physical modelling, specifically in fitting the diurnal tidal components particular to this station, while the data-based residuals can better adjust to observations, leading to a lower correlation (p=-0.01). The cases of SU2 and SU3 (p=0.03 and 0.02, respectively) are being investigated, but probable cause is also agreement with theoretical models due to local conditions. Higher correlation in other stations (p$\geq$0.11) indicates the methods of physical and data-based modelling might have a tendency to converge to similar results as compared to frequency filtering, having low correlation in all cases ($\abs{\mathrm{p}}<$0.03). The presence of the terdiurnal tide in the FIR residuals is a factor contributing to this result.

\subsection{Tidal removal and analysis of the system dynamics}\label{sec:nonlin}

\begin{figure}
    \centering
    \includegraphics[width=8.2cm]{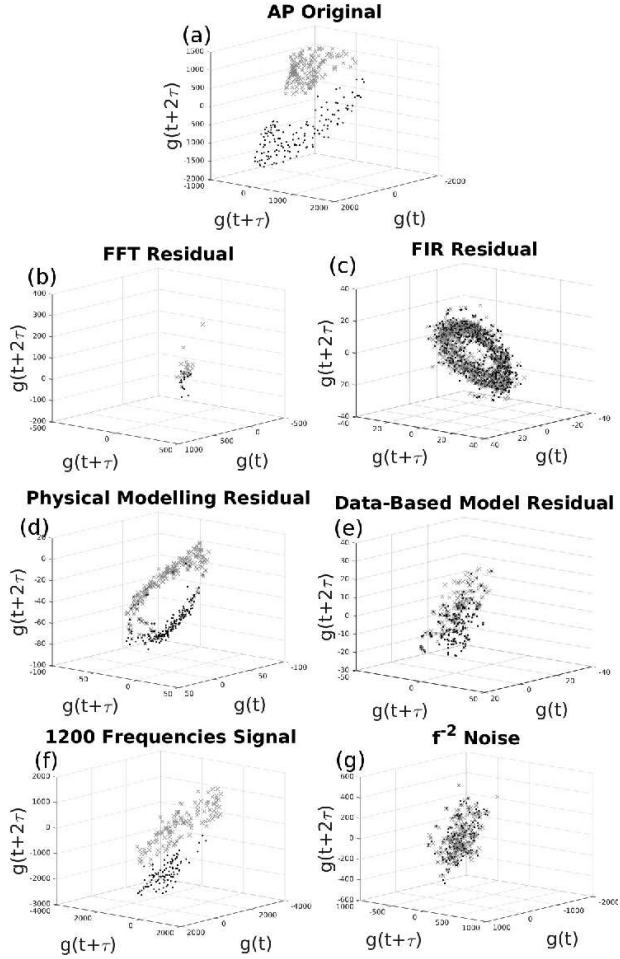}
    \caption{Map of the intersection of the trajectory with the surface $x_4=\langle x_4\rangle$ for the original gravity signal of AP station (a), gravity residuals obtained from the different methods (b-e), artificial signal generated by 1200 sinusoids corresponding to the tidal frequencies (f), and artificial red-noise (g). Similar patterns are observed for the other stations. The selected section is the one with fourth dimension equal to the mean fourth coordinate of the reconstructed attractor. Black dots are the intersections of the attractor trajectory with the surface in the direction of increasing $x_4$, and grey crosses are intersections in the direction of decreasing $x_4$. Units: nm s$^{-2}$.}
    \label{fig:poin}
\end{figure}

The removal of tides employing linear methods is an approximation, once nonlinearities are also present \cite{Munk1966}. It is important then to check for artificial changes in the overall system dynamics after applying these techniques. One fast approach is to observe the state-space plot $(x_1;x_2;...;x_m)=(g(t);g(t+\tau);...;g(t+(m-1)\tau))$ and compare it with known features. For this it is required to select optimal time-delay $\tau_{opt}$ and a dimension $m$ that permits maximal unfolding of the dynamical structure. This optimal time-delay is obtained by the time-lag for which the delayed average mutual information reaches its first minimum \cite{Abarbanel1996}, in the present case between 200 and 233 samples (3.3--3.9h). We will adopt $\tau_{opt}$=216 samples (3.6h) as a compromise to enable the comparison among stations, without significant disruption to the observation of the state-space plots. In order to minimize numerical errors, we choose the embedding dimension $m$ to be the minimum number of coordinates necessary completely describe the attractor from the data. This minimal value is calculated by the method of false-nearest neighbours\cite{Kantz1997}, i.e., by producing state-space plots with increasing embedding dimension $m'=$2,3,4,... and calculating the percentage of nearest neighbours to each point. When this percentage drops to zero the state-plot dimension is the minimum embedding dimension ($m=m'$). The value observed for the gravity time series from all stations is $m=4$. Comparatively, this value situates between what is expected from a periodic or a stochastic system. The value obtained of $m=4$ for the gravity system is also compatible with analogous observation of $m=4-6$ for shallow water ocean level \cite{Frison1999}. For visualization, a surface of section is arbitrarily selected as the hyperplane $x_4=\langle g(t+3\tau)\rangle$, and a 3-dimensional map is generated. This map is constituted by the locations where the trajectory crosses the surface section, and, additionally, we have colour-coded the direction of crossing. Adopting the same parameters for the original gravity series, the residuals were observed. The case for Apache Point time-series is shown in Fig. \ref{fig:poin}. The same result is observed on the other stations.

The original gravity series has a helicoidal structure on the map, with points uniformly distributed along its shape and a clear spatial division between the different directions of crossing (Fig. \ref{fig:poin}(a)). As reference, a sine wave (in principle a 2-dimensional system) would present a circular shape only when the chosen section section matches with the plane where lie all the points, but if this choice is not optimal, only two points (one for each direction of crossing) appear. Analogous effect is observed on the signal composed by 1200 sinusoids (Fig. \ref{fig:poin}(f)), where the distribution of points is approximately uniform in a stretched region and the spatial separation of different directions of crossing is also observed. The opposite case, the red-noise structure (Fig. \ref{fig:poin}(g)), as expected is not uniform and no spatial separation of crossing directions is present. Regarding the residuals from the different methods, the physical modelling (Fig. \ref{fig:poin}(d)) also shows a helicoidal structure, although in squeezed and modified form compared with the original time-series, and the separation of crossing directions. Data-based modelling hellicoidal features are less evident, with pattern approaching red-noise (Fig. \ref{fig:poin}(e)). The FIR residuals (Fig. \ref{fig:poin}(c)) show a torus instead, and FFT residuals (Fig. \ref{fig:poin}(b)) show isolated points, both without spatial distinction of crossing directions, meaning that the residuals from these two methods are fundamentally different from the dynamics of the original system. The difference in shape between the original gravity time-series and the time-series from 1200 sinusoidals generated from the Tamura tidal table (Fig. \ref{fig:poin}(f)) suggests that the underlying dynamics of the geophysical system is more complex.

The standard measure to determine if a system is predictable or chaotic is by the largest Lyapunov exponent ($\lambda_1$), defined as the exponential increase of distance between initial neighbours in the embedded space. As a general rule, a very large positive Lyapunov exponent indicates noise (stochasticity), a finite positive value is an indication of deterministic chaos, zero indicates limit-cycle stability (e.g. a sinusoid) and a negative value indicates point stability (convergence of signals). We have implemented an algorithm based on Rosenstein \cite{Rosenstein1993} method for the calculation of $\lambda_1$ by analysing how the nearest-neighbour to a base point in the embedded space will diverge after times $\tau$,2$\tau$,...,i$\tau$. The optimal delay of complete unfolding, $\tau_{opt}=216$ minutes, was too large, not producing exponential divergence/convergence, hence the embedding delay was adjusted empirically to 30 minutes for the Lyapunov exponent analysis. The larger delay is likely due to the low-period tidal frequencies, however not appropriate to capture the nonlinear features of the time-series. The average over 2000 randomly chosen initial base points was employed for improved statistics. The results obtained for the original time-series consisted of positive largest Lyapunov exponent (Table \ref{tb:lyapunov}). These are indicators for a chaotic nature of the signal. Additionally, the values suggest that the Earth responds to the tides by inheriting a small sensitivity to the initial conditions, thus enhancing the oscillations promoted by tides instead of damping them. The numbers obtained for gravity signals are consistent with observations of another time-series with tidal effects, the shallow water ocean levels \cite{Frison1999}, of $\lambda_1$=0.57--4.54 bits h$^{-1}$. Applying the same procedure for the residuals, the Lyapunov exponent values decreased slightly for each station, but the sensitivity to the initial conditions behaviour is maintained (Table \ref{tb:lyapunov}). As the largest Lyapunov exponent relates to the entropy of the signal, this reduction indicates that the residuals are less entropic than the original signal, as expected. Physical modelling residuals presented Lyapunov exponents closer to the original time-series, while other methods presented larger reductions but also greater differences between the stations (Table \ref{tb:lyapunov}). It is not possible to unambiguously infer that larger reductions are due to the removal of tides instead of other components which might be of geophysical interest. 

\begin{table}
\begin{minipage}{120mm}
\centering
\caption{Largest Lyapunov exponent [bits h$^{-1}$] of original gravity time-series and residuals after filtering for all stations}
\resizebox{120mm}{!}{
\begin{tabular}{cccccc}
Station & Original series &  FFT res. & FIR res.  & Phys. Model. res. &  Data-based res. \\ \hline
AP & $1.412\pm 0.007$ & $0.349\pm0.007$ &  $0.77\pm0.03$& $0.93\pm0.02$ & $0.63\pm0.02$\\
BF & $1.208\pm0.005$ & $0.29\pm0.01$ & $0.53\pm0.03$ & $0.84\pm0.01$ & $0.26\pm0.01$\\ 
MA & $1.317\pm0.006$ & $0.80\pm0.03$ & $0.76\pm0.02$ & $0.73\pm0.01$ & $0.42\pm0.02$\\ 
NY & $0.860\pm0.009$ & $0.35\pm0.04$ & $0.41\pm0.03$ & $0.81\pm0.02$ & $0.61\pm0.02$\\ 
SU & $1.326\pm0.007$ & $0.68\pm0.04$ & $0.58\pm0.03$ & $0.90\pm0.02$ & $0.84\pm0.02$\\ 
\end{tabular}}
\label{tb:lyapunov}
\end{minipage}
\end{table}

\subsection{Example of application of the residuals}\label{sec:appl}

\begin{figure}
    \centering
    \includegraphics[width=6cm]{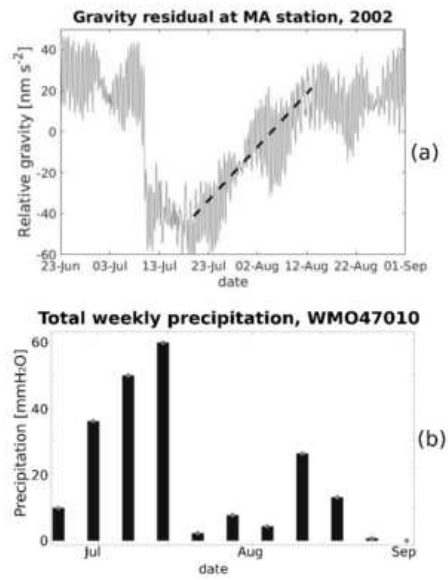}
    \caption{Observation of gravity drop in Matsushiro station associated with peak of precipitation and monthly drift associated to evapotranspiration and water in infiltration. The observation follows model in Imanishi et al (2013). Pluviometric data from nearby meteorological station in Nagano city, Japan.}
    \label{fig:hydro}
\end{figure}

The residuals from the physical and data-based modelling still appear to present some oscillatory behaviour in the time-domain, however their amplitudes have reduced considerably from the $2000\text{nm s}^{-2}$ daily levels to the $100 \text{nm s}^{-2}$ and $10\text{--}50\text{nm s}^{-2}$ daily levels, respectively. This reduction enables, for example, the observation of hydrology-induced gravity variations in the time-series. This has been object of previous study on the MA station (Matsushiro, Japan) by Imanishi et al. \cite{Imanishi2013}, who developed a model predicting a sharp drop of gravity measurement in the order of tenths of $\text{nm s}^{-2}$ during strong rainfall events, and a slow increase of gravity in the subsequent dry weeks, at rates associated to local evapotransporation and water infiltration phenomena. Figure \ref{fig:hydro} reveals that this behaviour is present in the residuals from data-based modelling residuals during the summer of 2002, where the gravity variations in the order of $60\text{nm s}^{-2}$ are associated with intense rainfall in the first weeks of July and the relative drier weeks in the period afterwards. This residuals presented daily oscillations around $20\text{nm s}^{-2}$ in this period, which allowed for the observation of the distinct features associated with the hydrological phenomena. The same could not be verified in the time-series of the residuals from physical modelling due to larger amplitude oscillations still present ($100\text{nm s}^{-2}$), masking the event. Co-seismic and post-seismic gravity changes still could not be observed at this resolution, yet the sensitivity level required ($0.1\text{--}50\text{nm s}^{-2}$) is very close. 

\section{Conclusions}

This work has presented three conceptually different approaches to filter tides in a general geophysical time-series, and classical implementations were applied to data from a network of superconducting gravimeters. These data sets in particular exhibit oscillations with amplitudes in the order of $2000\text{nm s}^{-2}$ of tidal origin, while, comparatively, the sensitivity of the instruments is of the order of $0.1\text{--}1\text{nm s}^{-2}$ and the phenomena of interest in the current frontier geophysical research is in the order of $0.1\text{--}100\text{nm s}^{-2}$. The three methods adopted here for removing the tides were frequency filtering, based on use of FFT or FIR filter to delete the undesired signal, physical modelling, where the undesired tidal signal was modelled from theoretical predictions and subtracted from the observation, and data-based modelling, where the parameters of the tides where obtained from the signal itself via least squares, and the reconstructed undesired signal was subtracted from the observation. The frequency filtering approach artificially distorted the signal (whether if implemented in frequency domain as FFT filtering or in time-domain as a FIR filter), and both the physical and data-based modelling methods, although not able to completely eliminate oscillations in the time-series, could reduce them significantly. In the time-domain, the frequency filtering residuals exhibited typically periodic/quasi-periodic daily amplitudes of the order of $10 \text{ nm s}^{-2}$ with FFT frequency filtering and $50 \text{ nm s}^{-2}$ with FIR frequency filtering. However, the frequency spectrum of the signal exhibited significant artificially induced changes, further confirmed by the change in the dynamics of the system. In comparison, the procedures of physical modelling and least-squares data-based modelling generated residuals with periodic/quasi-periodic daily amplitudes in the time-series typically in the order of $100 \text{ nm s}^{-2}$ and $10-50 \text{ nm s}^{-2}$, respectively, without significant disruption to the frequency domain (except reduction of the tidal peaks). Our dynamical analysis showed that the original system exhibits positive maximal Lyapunov exponent, an indicator of chaos, and this is preserved in the residuals. The data-based residuals reached enough sensitivity to monitor hydrology-related gravity changes in MA station (Matsushiro, Japan) in the order of $60 \text{ nm s}^{-2}$, and could also be applied to observation of any other phenomena above this level.

\end{doublespacing}

\section*{Supplementary Materials and Data Availability}
See the Supplementary Materials for the power spectral density plots of all the gravity residuals obtained from the eight time-series studied. IGETS network operators and GFZ-Potsdam provided the raw gravity data at http://isdc.gfz-potsdam.de/igets-data-base/.

\section*{Acknowledgments}
We thanks the participants of the 35th General Assembly of the European Seismological Commission for comments on preliminary results. The authors are grateful to all IGETS contributors, particularly to the station operators and to ISDC/GFZ-Potsdam for providing the original gravity data used in this study. We also thank the developers of ATLANTIDA3.1 and UTide. Part of this work was performed using the ICSMB High Performance Computing Cluster, University of Aberdeen. We also thanks M. Thiel and A. Moura for reviewing a preliminary version and making comments on the methods section and M.A. Ara\'ujo for comments on Lyapunov exponents.

Funding: A. Valencio is supported by CNPq, Brazil [206246/2014-5]; and received a travel grant from the School of Natural and Computing Sciences, University of Aberdeen [PO2073498], for a presentation including preliminary results.

\appendix
\section{Supplementary Materials}
On the Results section, Fig.4 provided the power spectral density of the original signal and the gravity residuals obtained from the different methods for the BF1 station, Schiltach/Black Forest, Germany. Figures (\ref{fig:psdAP}--\ref{fig:psdSU3}) here in the Supplementary Material present the resulting plots for the same analysis in the remaining stations.

Comparing the procedures for tidal removal, it should be noted in all cases the distortion of the spectrum after the procedures of frequency filtering. For FFT filtering, the frequency gaps related to removed tides are significant, implying in considerable bandwidths unavailable from geophysical analysis, and artificial loss of information for physical applications. The case of FIR filtering maintained for most stations the terdiurnal tides (around $3.5\cdot10^{-5}\text{ Hz}$) and the overall frequency spectrum usually present distortions most evident around $0.5\cdot10^{-5}\text{ Hz}$ and $4.4\cdot10^{-5}\text{ Hz}$. The considerable peaks in diurnal (around $1.16\cdot10^{-5}\text{ Hz}$) and especially semidiurnal (around $2.31\cdot10^{-5}\text{ Hz}$) bands in the residuals from physical modelling indicate present-day challenges in the theoretical description of the Earth system. These residuals can vary considerably with location, depending on how well the theoretical model reflects the local environment. Particular challenges are shown at the SU stations (Sutherland, South Africa). The data-based residuals exhibits a greater reduction of tidal amplitudes compared to physical modelling, although tides remain present. The background noise level is being lowered in amplitude in the modelling procedures compared to the filtering methods. As a results it is possible to observe the quaterdiurnal tides (around $4.63\cdot10^{-5}\text{ Hz}$) in most stations using the modelling methods.

\begin{figure}
    \centering
    \includegraphics[height=18cm]{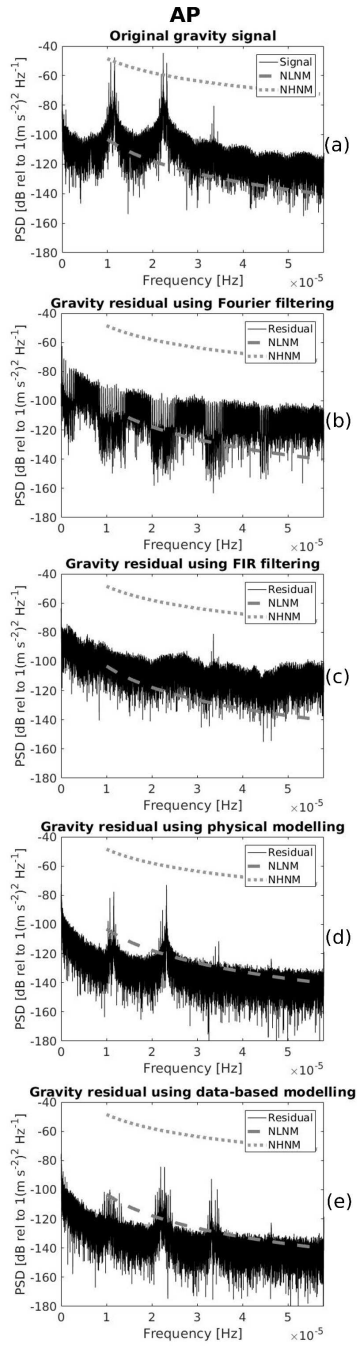}
    \caption{Lomb-Scragle power spectral density of the original gravity signal (a) and gravity residuals (b-e) from AP instrument, Apache Point (New Mexico). New Earth Low Noise (NLNM) and High Noise Models (NHNM) are indicated.}
    \label{fig:psdAP}
\end{figure}
\begin{figure}
    \centering
    \includegraphics[height=18cm]{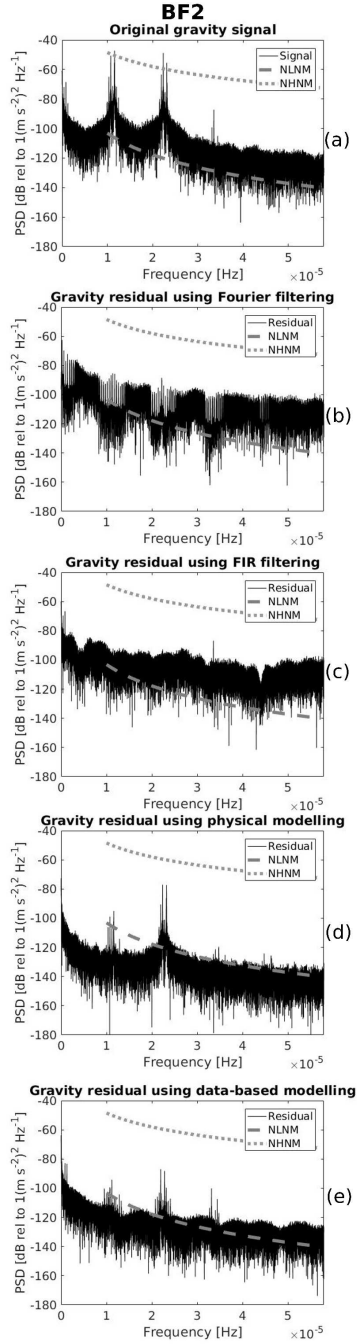}
    \caption{Lomb-Scragle power spectral density of the original gravity signal (a) and gravity residuals (b-e) from BF2 instrument, Schiltach/Black Forest (Germany). New Earth Low Noise (NLNM) and High Noise Models (NHNM) are indicated.}
    \label{fig:psdBF2}
\end{figure}
\begin{figure}
    \centering
    \includegraphics[height=18cm]{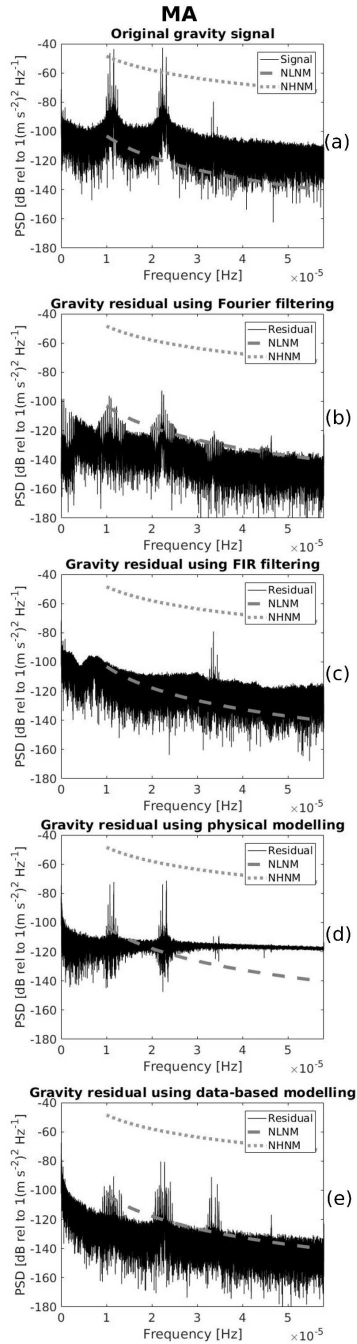}
    \caption{Lomb-Scragle power spectral density of the original gravity signal (a) and gravity residuals (b-e) from MA instrument, Matsushiro (Japan). New Earth Low Noise (NLNM) and High Noise Models (NHNM) are indicated. The sharpening of the noise spectrum above $2.5\cdot10^{-5}$ Hz in (d) is an artificial effect of the Lomb computation. Only the presence of peaks should be considered in this case.}
    \label{fig:psdMA}
\end{figure}
\begin{figure}
    \centering
    \includegraphics[height=18cm]{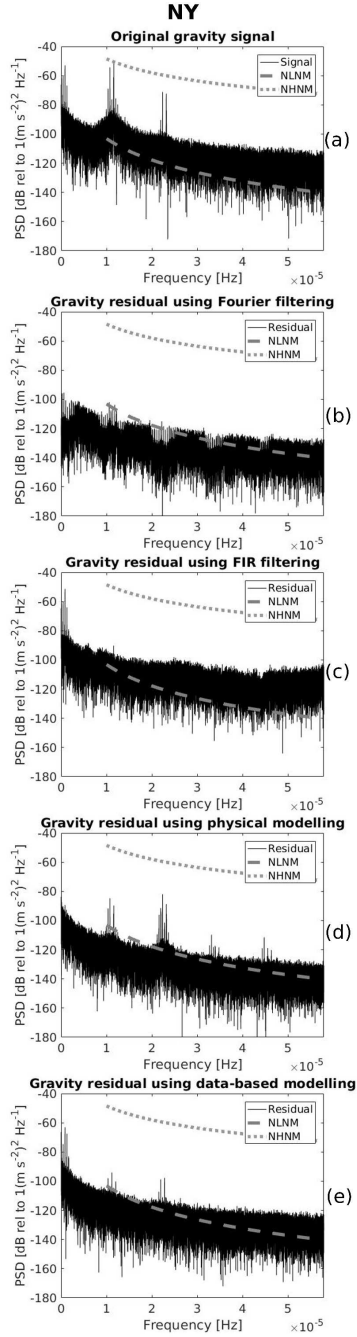}
    \caption{Lomb-Scragle power spectral density of the original gravity signal (a) and gravity residuals (b-e) from NY instrument, Ny-\AA lesund (Svalbard island, Norway). New Earth Low Noise (NLNM) and High Noise Models (NHNM) are indicated.}
    \label{fig:psdNY}
\end{figure}
\begin{figure}
    \centering
    \includegraphics[height=18cm]{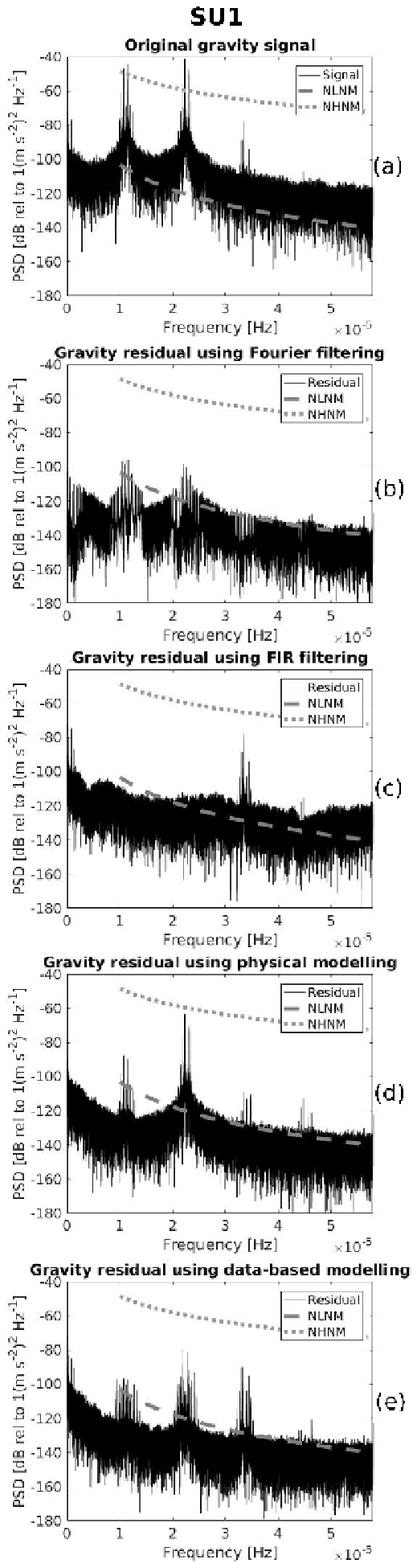}
    \caption{Lomb-Scragle power spectral density of the original gravity signal (a) and gravity residuals (b-e) from SU1 instrument, Sutherland (South Africa). New Earth Low Noise (NLNM) and High Noise Models (NHNM) are indicated.}
    \label{fig:psdSU1}
\end{figure}
\begin{figure}
    \centering
    \includegraphics[height=18cm]{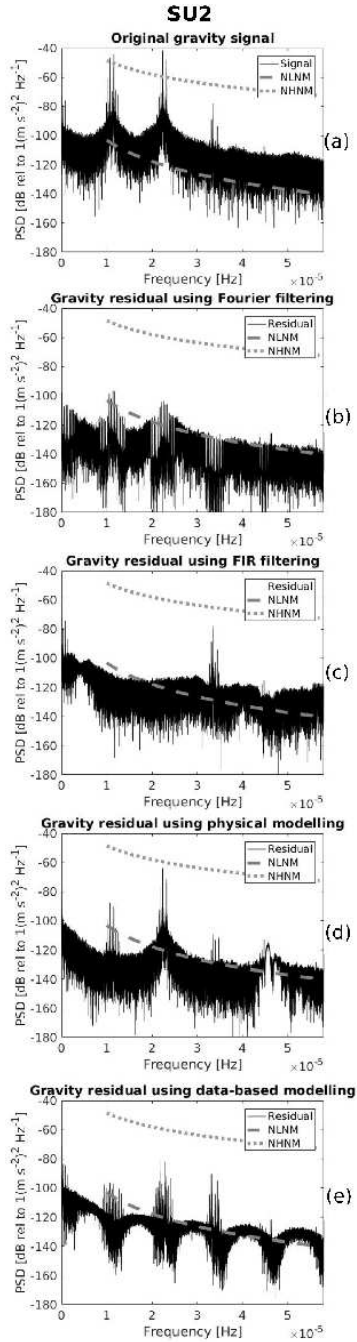}
    \caption{Lomb-Scragle power spectral density of the original gravity signal (a) and gravity residuals (b-e) from SU2 instrument, Sutherland (South Africa). New Earth Low Noise (NLNM) and High Noise Models (NHNM) are indicated. The fluctuations in the background noise in (e) are artificial effects of the Lomb computation, analogous to the windowing effects in regular periodograms. Only the presence of peaks should be considered in this case.}
    \label{fig:psdSU2}
\end{figure}
\begin{figure}
    \centering
    \includegraphics[height=18cm]{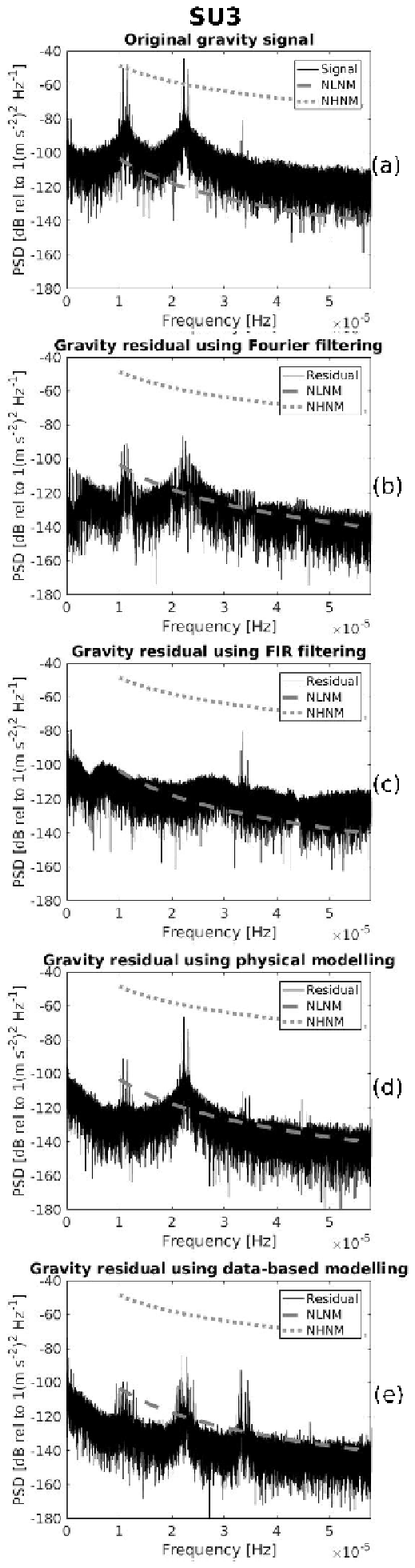}
    \caption{Lomb-Scragle power spectral density of the original gravity signal (a) and gravity residuals (b-e) from SU3 instrument, Sutherland (South Africa). New Earth Low Noise (NLNM) and High Noise Models (NHNM) are indicated.}
    \label{fig:psdSU3}
\end{figure}

\end{document}